\documentclass[aps,prl,preprint,superscriptaddress,showpacs]{revtex4}

\usepackage{graphicx}

\begin{document}

\title{A Non-axisymmetric Magnetorotational Instability
of a Purely Toroidal Magnetic Field}

\author{Rainer Hollerbach}
\address{Department of Applied Mathematics, University of Leeds,
Leeds, LS2 9JT, United Kingdom}
\author{G\"unther R\"udiger, Manfred Schultz, D. Elstner}
\address{Astrophysikalisches Institut Potsdam, An der Sternwarte 16,
D-14482 Potsdam, Germany}

\date{\today}

\begin{abstract}
We consider the flow of an electrically conducting fluid between
differentially rotating cylinders, in the presence of an externally
imposed toroidal field $B_0(r_i/r)\,{\bf\hat e}_\phi$.  It is known
that the classical, axisymmetric magnetorotational instability does
not exist for such a purely toroidal imposed field.  We show here
that a non-axisymmetric magnetorotational instability does exist,
having properties very similar to the axisymmetric magnetorotational
instability in the presence of an axial field.
\end{abstract}


\maketitle

The magnetorotational instability (MRI) is a mechanism whereby a
hydrodynamically stable differential rotation flow may be destabilized by
the addition of a magnetic field.  An obvious question then is whether
different orientations of the field yield different types of instability,
or the same, or none at all.  The original view was that the axial
component of the field is the only important one, with any azimuthal
component playing no essential role, and incapable of producing any
instabilities on its own \cite{V59, BH91}.  This view was altered
by the discovery that a mixed axial and azimuthal field yields
instabilities quite different in many ways from those found with a purely
axial field \cite{HR05, RH05}.  In this letter we show that even a purely
azimuthal field yields an MRI, and compare its properties with the
previously known types.

While its most important application is to astrophysical accretion disks
\cite{BH91}, the MRI was originally discovered in the much simpler
Taylor-Couette problem, consisting of the flow between differentially
rotating cylinders \cite{V59}.  Because of its relative simplicity, this
geometry has proven particularly amenable both to theoretical analyses of
the MRI \cite{Ji,RZ,RSS}, as well as to its recent experimental realization
\cite{S06, R06}.  We too shall examine the MRI in this context.

Consider therefore an
electrically conducting fluid confined between two concentric cylinders of
radii $r_i$ and $r_o$, rotating at rates $\Omega_i$ and $\Omega_o$, chosen
to satisfy $\Omega_o r_o^2>\Omega_i r_i^2$.  That is, the angular momentum
increases outward, so by the familiar Rayleigh criterion the flow is
hydrodynamically stable, with the angular velocity given by
$$\Omega(r)=A + B/r^2,\eqno(1)$$
where
$$A=\frac{\Omega_or_o^2-\Omega_ir_i^2}{r_o^2-r_i^2},\quad
  B=\frac{r_i^2r_o^2(\Omega_i-\Omega_o)}{r_o^2-r_i^2}.\eqno(2)$$
In this work we will fix $r_o=2r_i$ and $\Omega_o=\Omega_i/2$,
so $A$ and $B$ simplify to $\frac{1}{3}\Omega_i$ and $\frac{2}{3}\Omega_i
r_i^2$, respectively.  The essence of the MRI then is to ask whether the
addition of a magnetic field can destabilize this flow.

If the imposed field is purely axial, ${\bf B}_0=B_0\,{\bf\hat e}_z$, the
profile (1) can be destabilized, provided the rotation rates are
sufficiently great, and the field strength $B_0$ is neither too weak nor
too strong \cite{Ji,RZ}.  Specifically, the magnetic Reynolds number ${\rm Rm}
=\Omega_ir_i^2/\eta$, where $\eta$ is the magnetic diffusivity, must exceed
$O(10)$, and the Lundquist number ${\rm S}=B_0r_i/\eta\sqrt{\rho\mu}$, where
$\rho$ is the density and $\mu$ the permeability, must be around $3-10$.

In contrast, if the imposed field is mixed axial and azimuthal, ${\bf B}_0
=B_0\,{\bf\hat e}_z + \beta B_0(r_i/r)\,{\bf\hat e}_\phi$, where $\beta$
is around $1-10$, then the profile (1) can again be destabilized, but at
very different rotation rates and field strengths \cite{HR05,RH05}.  The
relevant parameter measuring the rotation rates is now not the magnetic
Reynolds number, but rather the hydrodynamic Reynolds number ${\rm Re}=
\Omega_ir_i^2/\nu$, where $\nu$ is the viscosity.  Similarly, the parameter
measuring the field strength is not the Lundquist number, but instead the
Hartmann number ${\rm Ha}=B_or_i/\sqrt{\rho\mu\eta\nu}$.  The MRI sets
in when ${\rm Re}\gtrsim O(10^3)$, and ${\rm Ha}\approx O(10)$.

To compare these results, we note that the two sets of parameters are
related by ${\rm Re=Rm\,Pm}^{-1}$ and ${\rm Ha=S\,Pm}^{-1/2}$, where
${\rm Pm}=\nu/\eta$ is the magnetic Prandtl number, a material property of the
fluid.  Typical values for liquid metals are $O(10^{-6})$.  Translating the
results for the purely axial field, we thus obtain ${\rm Re}\gtrsim O(10^7)$
and ${\rm Ha}\approx O(10^4)$, both several orders of magnitude greater than
for the mixed field.  It is perhaps not surprising then that the MRI has
been obtained experimentally for the mixed field \cite{S06,R06}, but not
(yet) for the purely axial field \cite{Ji}.

As different as they are, one feature these two types of MRI have in
common is that they are both axisymmetric.  Non-axisymmetric modes have
also been explored, for both the purely axial \cite{RSS} as well as the
mixed fields \cite{RH05}.  For a purely axial field the relevant
parameters are still $\rm Rm$ and $\rm S$, but the critical values $\rm Rm_c$
are somewhat larger than for the axisymmetric modes, indicating that the
axisymmetric MRI is the most unstable mode.  For mixed fields, one finds
--- perhaps somewhat surprisingly --- that adding an azimuthal component
now has minimal effect, certainly far less than the reduction by four
orders of magnitude found for the axisymmetric modes.  Evidently the
relevant parameters continue to be $\rm Rm$ and $\rm S$, rather than
$\rm Re$ and $\rm Ha$.

What we wish to show in this letter then is that for these
non-axisymmetric modes, one can impose a purely azimuthal field,
$B_0(r_i/r)\,{\bf\hat e}_\phi$, and still obtain an MRI, having all the
characteristics of the previous non-axisymmetric types of MRI \cite{RSS,
RH05}.  To this end, we solve the linear stability equations
$${\rm Rm}\,\frac{\partial{\bf b}}{\partial t}=\nabla^2{\bf b}
 +\nabla\times({\bf u\times B}_0)
+ {\rm Rm}\,\nabla\times({\bf U}_0\times{\bf b}),\eqno(3)$$
$${\rm Re}\,\frac{\partial{\bf u}}{\partial t}=-\nabla p + \nabla^2{\bf u}
 + {\rm Ha}^2\,(\nabla\times{\bf b})\times{\bf B}_0
  +{\rm Re}\,({\bf U}_0\times\nabla\times{\bf u}
   +{\bf u\times\nabla\times U}_0),\eqno(4)$$
where ${\bf U}_0=r\Omega(r)\,{\bf\hat e}_\phi$ is the profile (1) whose
stability we are exploring, and ${\bf B}_0=B_0(r_i/r)\,{\bf\hat e}_\phi$
is the imposed azimuthal field.  Length has been scaled by $r_i$, time by
$\Omega_i^{-1}$, ${\bf U}_0$ and $\bf u$ as $r_i\Omega_i$, ${\bf B}_0$ as
$B_0$, and $\bf b$ as ${\rm Rm}B_0$.

Taking the $t$, $z$ and $\phi$ dependence to be of the form
$\exp(\sigma t+ikz+im\phi)$, and using also $\nabla\cdot{\bf b}=0$ to
eliminate $b_z$, the $r$ and $\phi$ components of (3) become
$${\rm Rm}\,\sigma b_r=\nabla^2 b_r - r^{-2}b_r
 - 2imr^{-2}b_\phi+imr^{-2}u_r
-{\rm Rm}\,im\Omega\,b_r,\eqno(5)$$
$${\rm Rm}\,\sigma b_\phi=\nabla^2 b_\phi - r^{-2}b_\phi + 2imr^{-2}b_r
                                        +imr^{-2}u_\phi
+2r^{-2}u_r-{\rm Rm}\,im\Omega\,b_\phi + {\rm Rm}\,\Omega'r b_r,
\eqno(6)$$
where primes denote $d/dr$.  The components of (4) have a similar structure,
but we will not need to refer to them in the subsequent discussion, and
hence do not list them.  The boundary conditions associated with (3) and (4)
are
$$b_r=b_\phi'+r^{-1}b_\phi=u_r=u_\phi=u_z=0\eqno(7)$$
at $r=r_i$ and $r_o$, corresponding to perfectly conducting, no-slip walls.
The resulting one-dimensional linear eigenvalue problem is solved by finite
differencing in $r$, as in \cite{RSS}.

Figure 1 shows stability curves for $m=1$, the only mode
that appears to become unstable.  We see how an MRI exists that is remarkably
similar to some of the results described above.  In particular, as ${\rm Pm}
\to0$, the relevant parameters are clearly once again $\rm Rm$ and $\rm S$,
with the MRI arising if ${\rm Rm}\gtrsim80$, and ${\rm S}\approx40$ yielding
the lowest value of $\rm Rm_c$.  The specific numbers are roughly an order of
magnitude greater than for the axisymmetric MRI in the axial field, but the
basic scalings, and even the detailed shape of the instability curves, are
identical.

\begin{figure}[htb]
\includegraphics[scale=0.5]{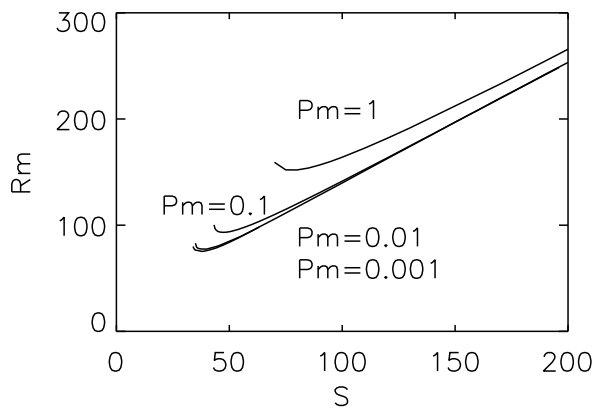}
\caption{The critical magnetic Reynolds number for the onset of the MRI,
as a function of the Lundquist number, for the different values of $\rm Pm$
indicated.  $m=1$, $r_o/r_i=2$, $\Omega_o/\Omega_i=0.5$.}
\end{figure}

Figure 2 shows the real and imaginary parts of $\sigma$ in the unstable
regime.  Remembering that time has been scaled by $\Omega_i^{-1}$, we see
that we obtain growth rates as large as $0.05\Omega_i$.  So again, while the
particular number 0.05 is about an order of magnitude smaller than for the
axisymmetric MRI in the axial field, this non-axisymmetric MRI is
clearly also growing on the basic rotational timescale.

To understand why this non-axisymmetric MRI exists even for a purely
toroidal field ${\bf B}_0$, for which it is known that the axisymmetric MRI
fails \cite{V59,BH91}, we need to consider the details of (5) and (6).  In
particular, note that for $m=0$, $b_r$ completely decouples from everything
else, and inevitably decays away.  Without $b_r$ though, the MRI cannot
proceed, as it relies on the term ${\rm Rm}\,\Omega'r b_r$ in (6).  In
contrast, for $m=1$, $b_r$ is coupled both to $b_\phi$, coming from
$\nabla^2{\bf b}$, and to $u_r$, from $\nabla\times({\bf u\times B}_0)$.
And once $b_r$ is coupled to the rest of the problem, the term
${\rm Rm}\,\Omega'r b_r$ then allows the MRI to develop.  Figure 3 presents
an example of these solutions, indicating how all three components of both
$\bf u$ and $\bf b$ are indeed present.

\begin{figure}
\vbox{
\includegraphics[scale=0.45]{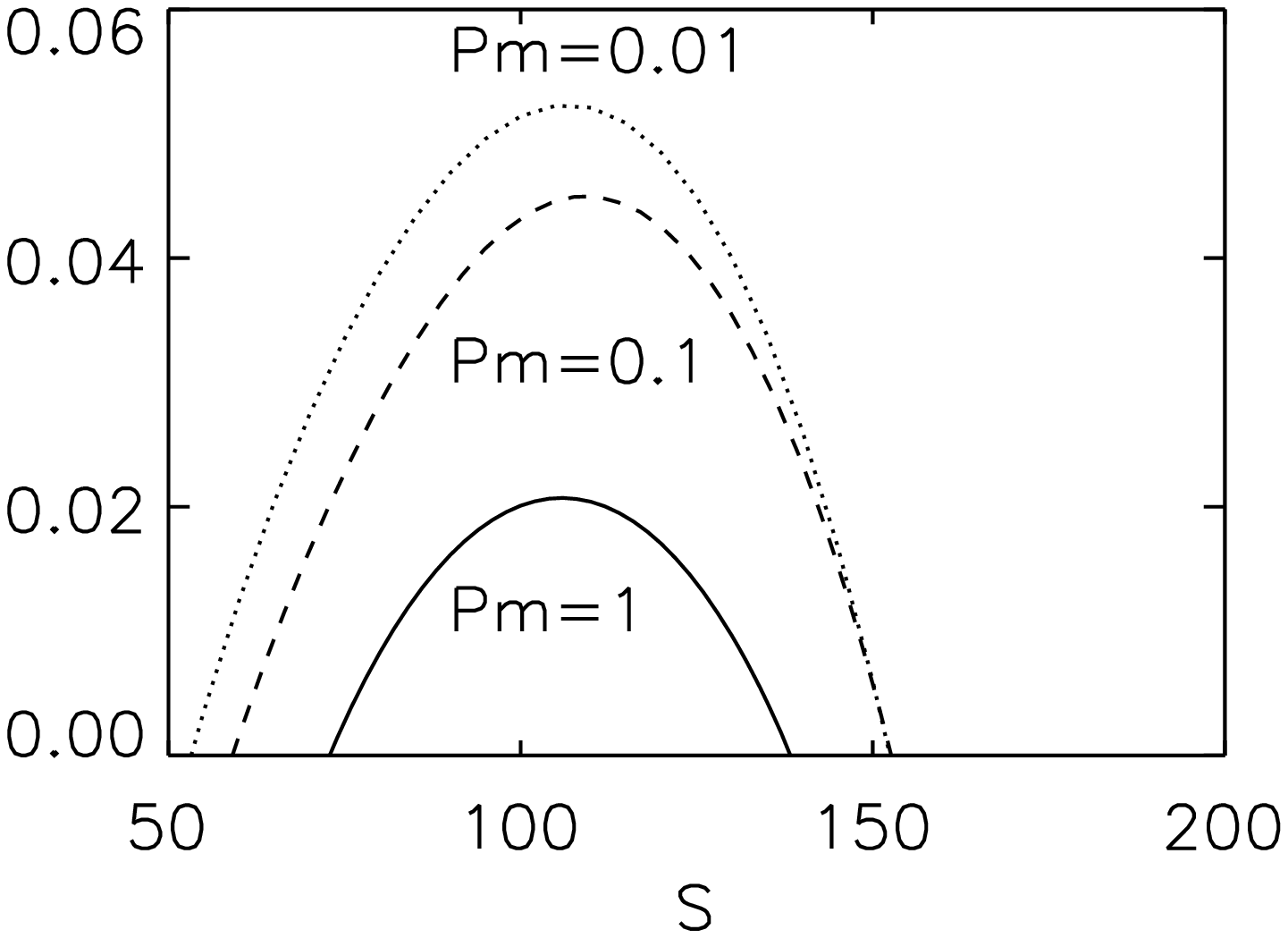}
\includegraphics[scale=0.45]{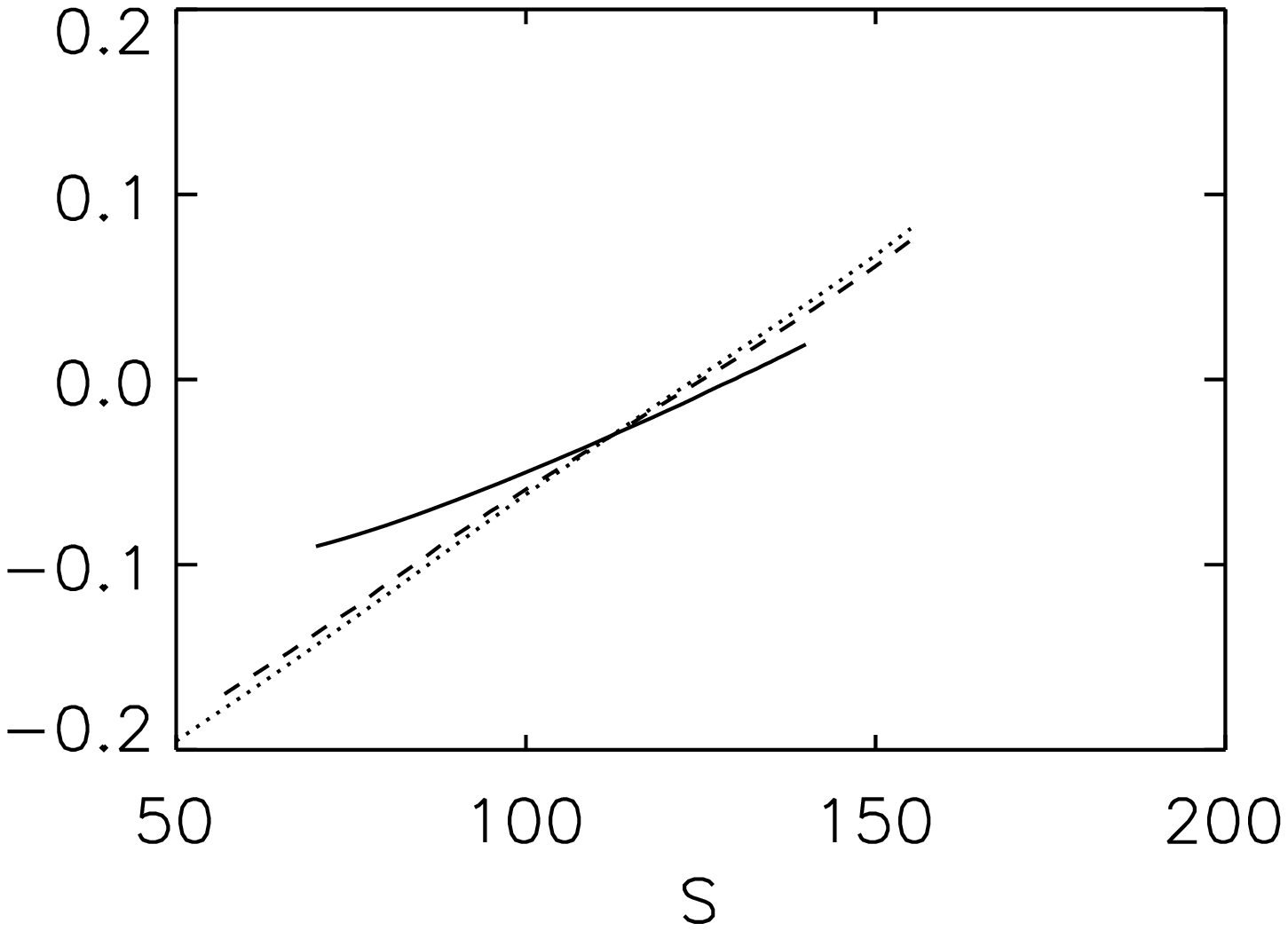}}
\caption{The real (top) and imaginary (bottom) parts of $\sigma$, as
functions of $\rm S$, with $\rm Rm$ fixed at 200.  The real part corresponds
to the growth rate, the imaginary part to the azimuthal drift rate.}
\end{figure}

\begin{figure}
\vbox{
\hbox{
\includegraphics[width=2.8cm,height=5.0cm]{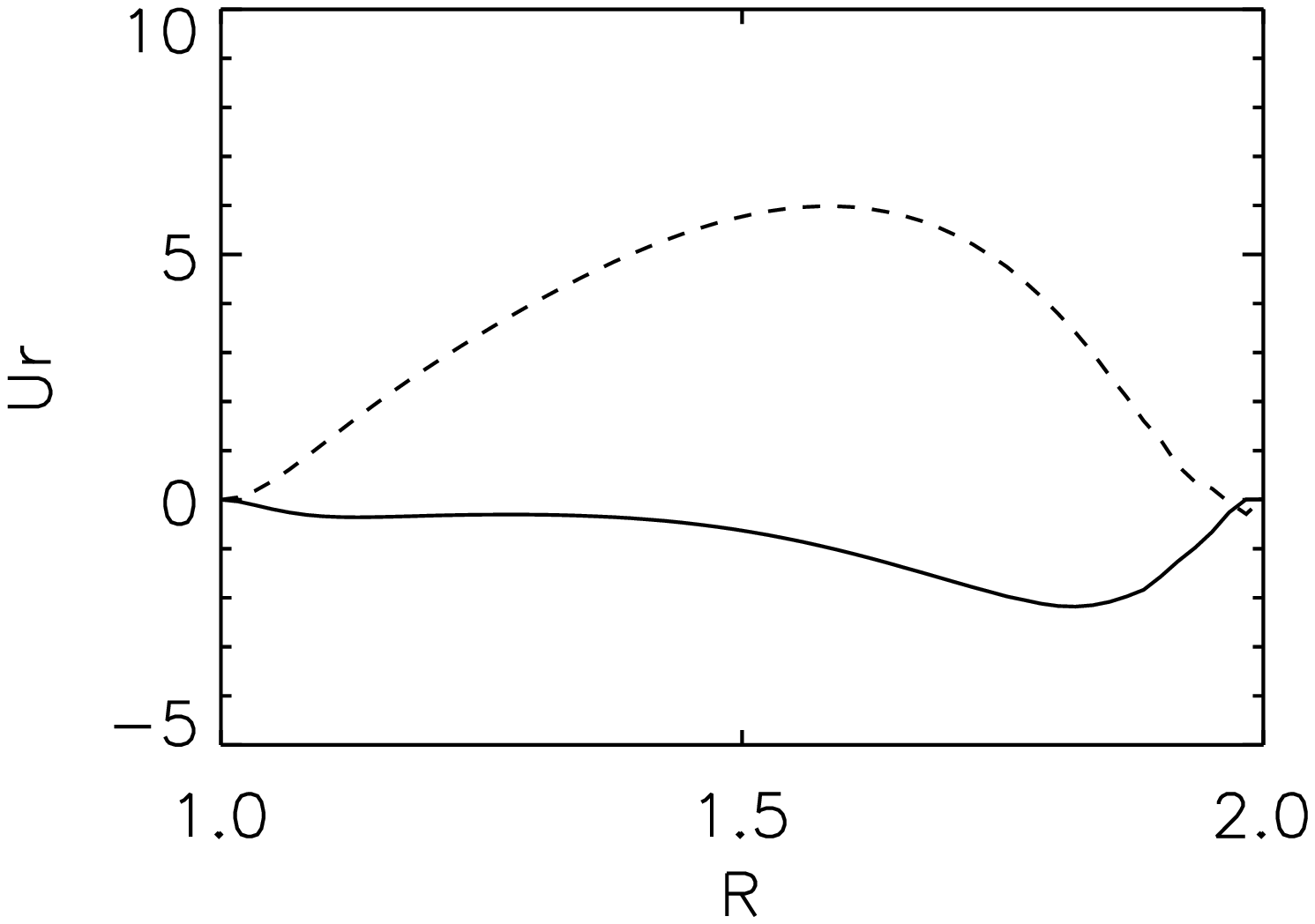}
\includegraphics[width=2.8cm,height=5.0cm]{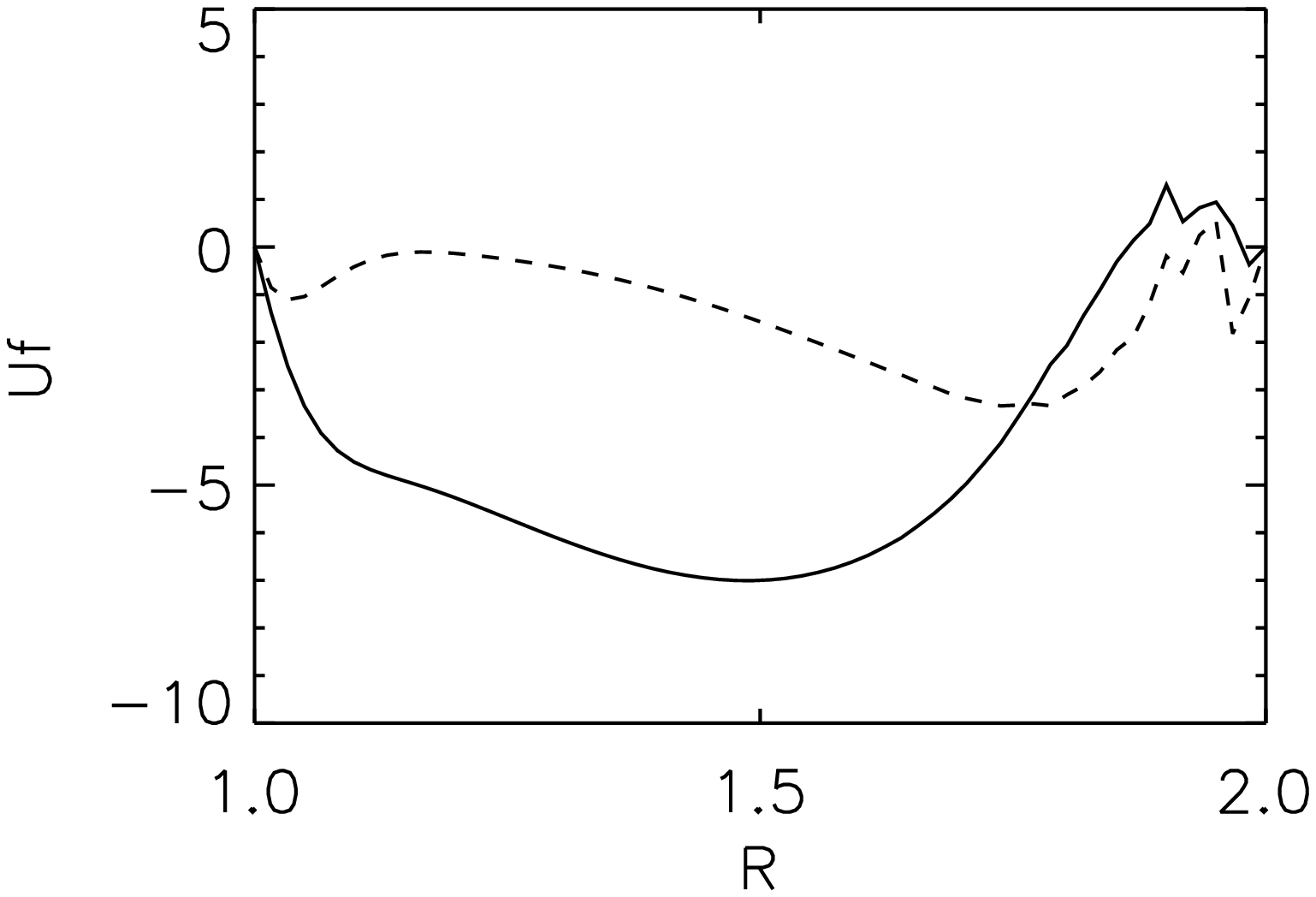}
\includegraphics[width=2.8cm,height=5.0cm]{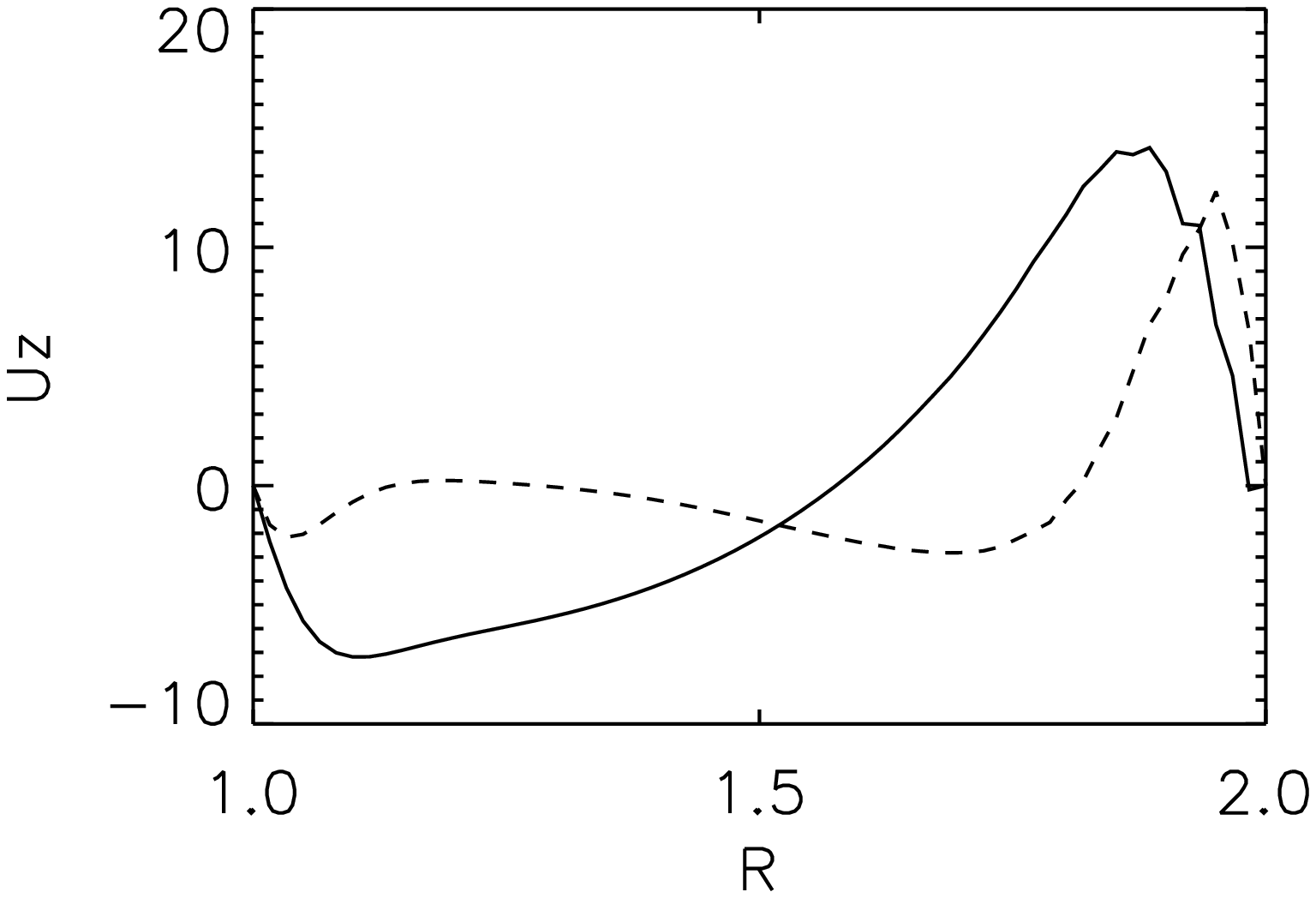}}
\hbox{
\includegraphics[width=2.8cm,height=5.0cm]{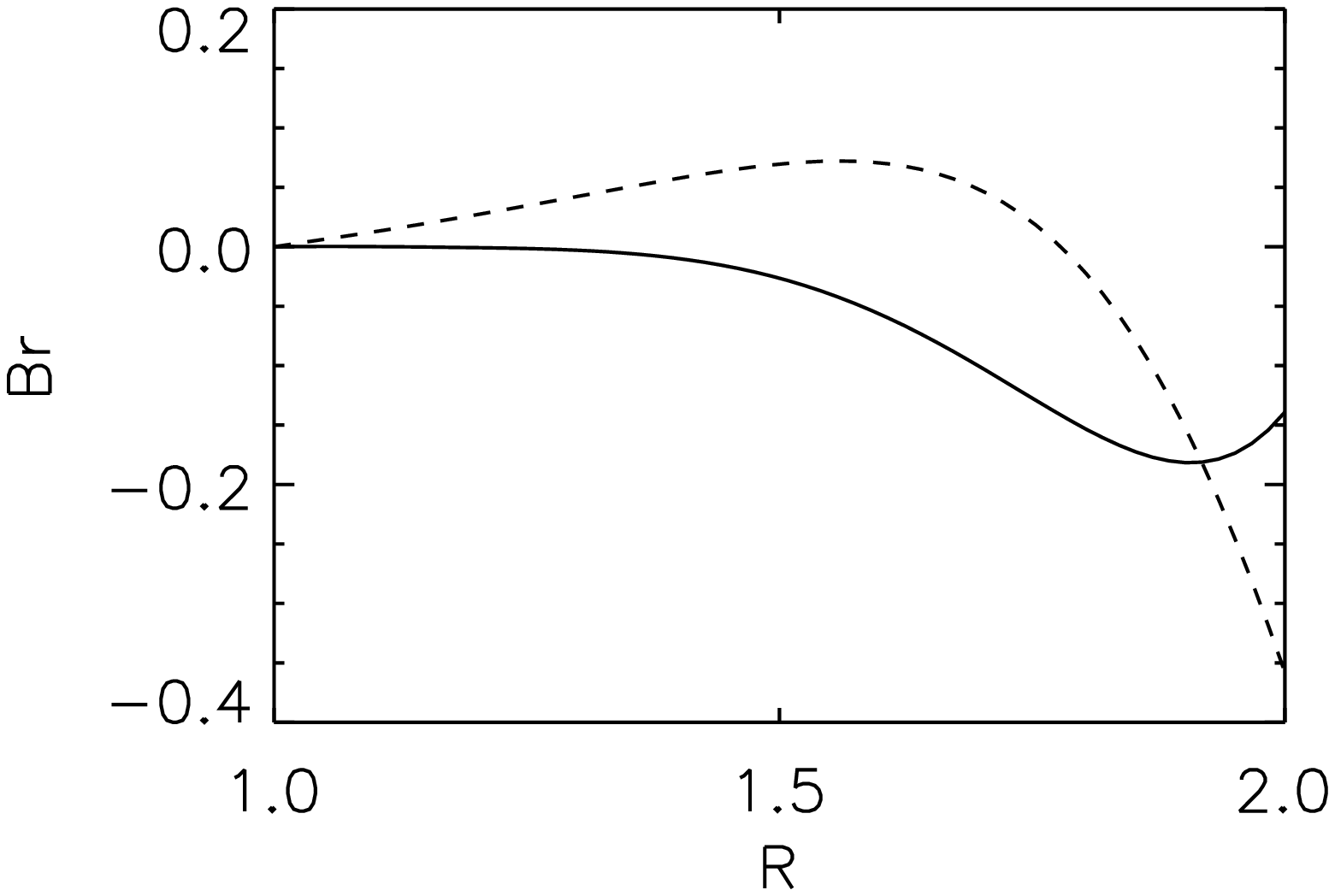}
\includegraphics[width=2.8cm,height=5.0cm]{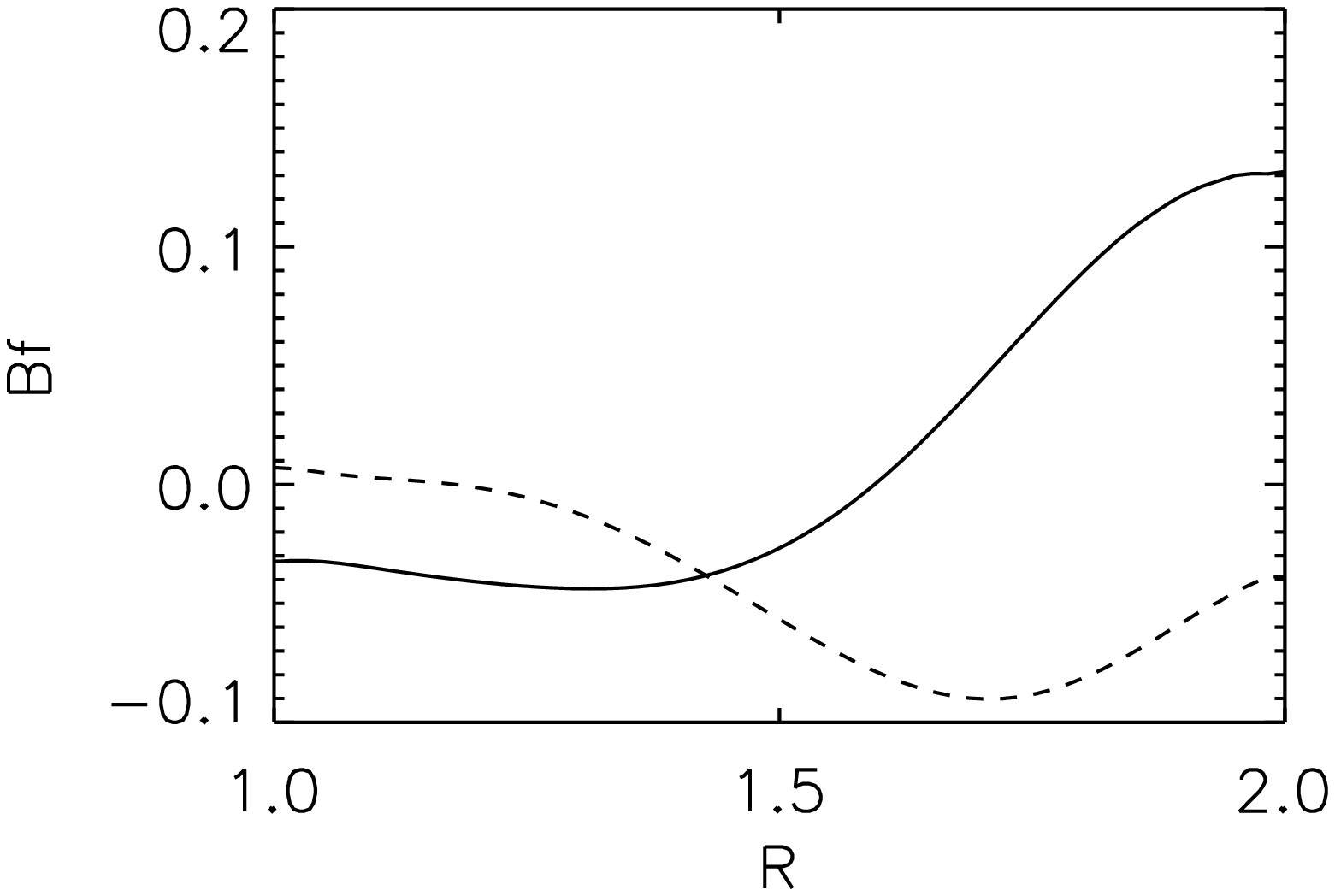}
\includegraphics[width=2.8cm,height=5.0cm]{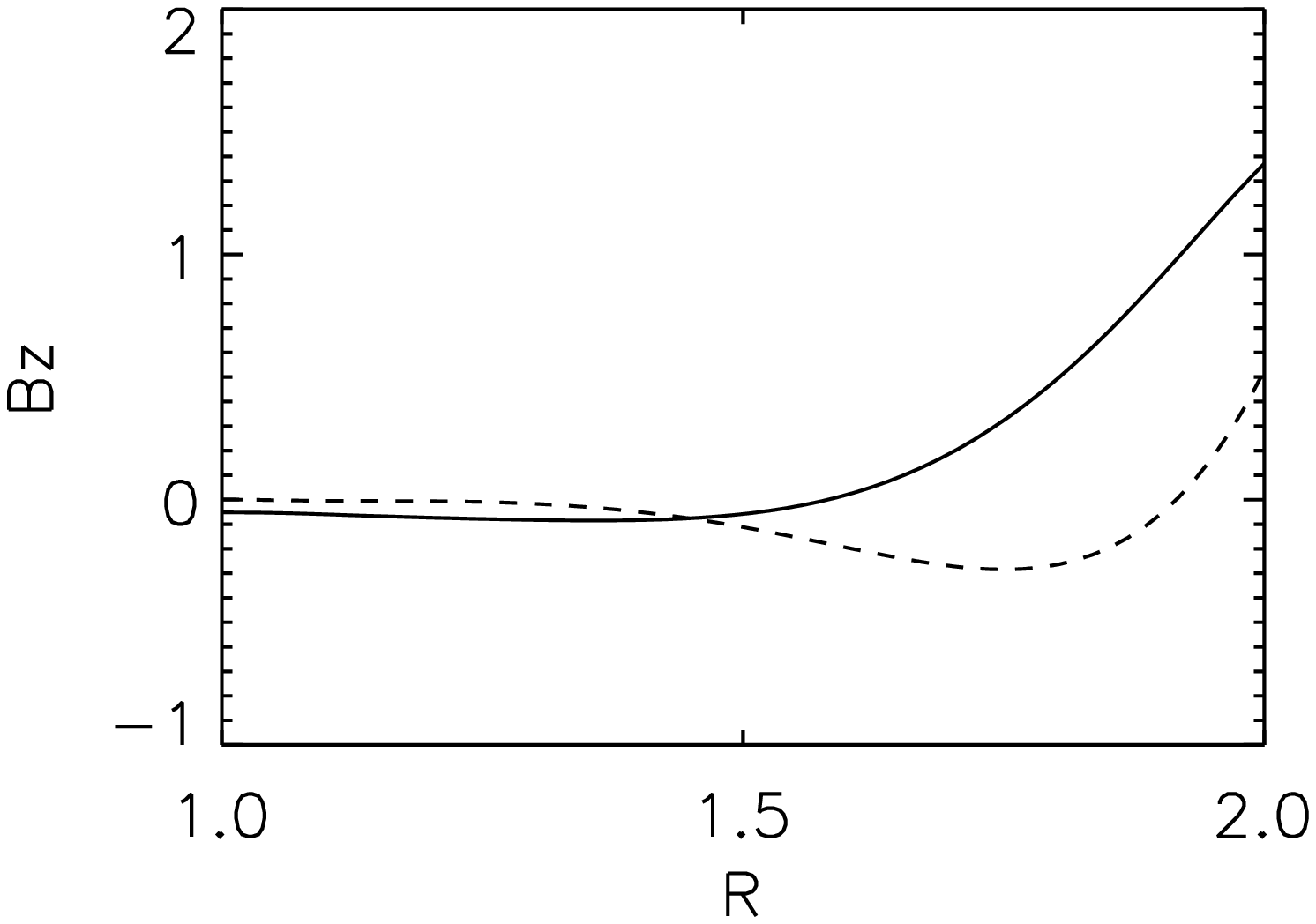}}
}
\caption{The marginally stable solution at ${\rm Pm}=0.1$, ${\rm S}=47.4$
and ${\rm Rm}=93$ (see Fig.\ 1).  From left to right the $r$, $\phi$ and
$z$ components of $\bf u$ (top) and $\bf b$ (bottom).  The real
parts are solid, imaginary parts dashed.  Note how both $\bf u$ and $\bf b$
have the $z$ components largest.  The azimuthal wavenumber is $k=1.88$.}
\end{figure}

There is however one aspect that is not entirely clear from this
analysis, namely why this non-axisymmetric MRI actually requires the term
${\rm Rm}\,\Omega'r b_r$ at all.  In particular, the axisymmetric MRI with
a mixed field does {\it not} rely on it, but instead on the term $2r^{-2}u_r$,
coming from $\nabla\times({\bf u\times B}_0)$ rather than ${\rm Rm}\,\nabla
\times({\bf U}_0\times{\bf b})$ \cite{HR05}.  The non-axisymmetric MRI is 
evidently more like the axisymmetric MRI with a purely axial field, which
also requires the term ${\rm Rm}\,\Omega'r b_r$, because in that case the
term $2r^{-2}u_r$ is absent.

To summarize then, we have shown that even if the externally imposed magnetic
field is purely toroidal, one still obtains an MRI, simply non-axisymmetric
rather than axisymmetric.  In other respects though this new instability is
remarkably like the classical axisymmetric MRI with a purely axial field,
indeed far more like it than the axisymmetric MRI with a mixed field, which
yielded fundamentally different scalings.  In contrast, the parameters here
continue to be $\rm Rm$ and $\rm S$, just as in the classical MRI.

Note though that attempting to obtain this non-axisymmetric MRI in a
laboratory experiment would be even more difficult than attempting to obtain
the classical axisymmetric MRI in an axial field.  First, the required
rotation rates would be even greater, ${\rm Re}\gtrsim O(10^8)$, with all
the difficulties that entails \cite{HF}.  Even more daunting, imposing an
azimuthal field of the required strength, ${\rm Ha}\approx O(10^5)$, would
require a current along the central axis in excess of $10^6$ A, surely far
beyond any feasible experiment.

This new non-axisymmetric MRI could have astrophysical applications though,
since many astrophysical objects do have predominantly azimuthal fields, in
which case the results presented here suggest that this non-axisymmetric
MRI could be preferred over the axisymmetric MRI.

Despite the similarities in their fundamental scalings, the fact that the
classical MRI is axisymmetric, whereas this new MRI is non-axisymmetric,
means the nonlinear equilibration, and hence the associated angular momentum
transport, are potentially quite different, which could again have
astrophysical implications.  Work on this is currently in progress.

Finally, one might ask how the results presented here change if one
allows for a more general toroidal field profile, ${\bf B}_0=(c_1r^{-1}
+c_2r)\,{\bf\hat e}_\phi$, where the term $c_2r$ corresponds to
an electric current flowing through the fluid itself, and not just along
the central axis.  Eq.\ (4) then contains an additional term
${\rm Ha}^2(\nabla\times{\bf B}_0)\times{\bf b}$, which opens up the
possibility of instabilities driven entirely by this current
$\nabla\times{\bf B}_0$, without any rotation necessarily present at all
\cite{Tay}.  Understanding how these current driven instabilities (also
$m=1$) interact with the magnetically catalyzed but ultimately
rotationally driven MRI presented here is also in progress.

\begin{acknowledgments}
This work was supported by the German Leibniz Gemeinschaft, under program
SAW.
\end{acknowledgments}

\end{document}